\documentclass[twocolumn,pra,showpacs]{revtex4}
\usepackage{amsmath,amsfonts}
\begin{document}
\draft
\title{\bf Screening of  electric field and nuclear EDM in non-stationary states of atoms and molecules }
\author{V.V. Flambaum
} 
\affiliation{
School of Physics, University of New South Wales,  Sydney 2052,  Australia}
\date{\today}

\begin{abstract}
According to the Schiff theorem, an external electric field vanishes at atomic nucleus in a neutral atom in a stationary state, i.e. it is completely shielded
by electrons. This makes a nuclear electric dipole moment (EDM)  unobservable. We show that if atom or molecule is not  in a stationary state (e.g. in a superposition of two stationary states), electric field on the nucleus is not zero and interaction with nuclear EDM does not vanish. In molecules this effect is enhanced by the ratio of nuclear mass to to electron mass, $M_n/m_e$ , since  nuclei in a molecule are slow (compare to electrons) and do not provide efficient screening in a non-stationary environment. Electric field on the nucleus may also affect nuclear reactions. 

 \end{abstract}

\maketitle
\section{Introduction}
Existence of electric dipole moments (EDM) of elementary particles, nuclei, atoms and molecules in a state with a definite angular momentum violates time reversal invariance (T) and parity (P). EDM also violates CP invariance if the CPT invariance holds. A very extensive experimental and theoretical activity related to EDM is motivated by the need to test unification theories predicting T, P, and CP violation. 


 However, there is a problem here.   A homogeneous static electric field does not accelerate neutral atom. This means that the total electric field ${\bf E}$ acting on the atomic nucleus is zero  since otherwise the charged nucleus would be accelerating, i.e. the external  field is completely shielded by atomic electrons.  The absence of the electric field means that the nuclear EDM $d$ is unobservable, ${\bf d \cdot E}=0$. One may also present this result differently:  
 total  atomic EDM is zero even if the nucleus has EDM, i.e. EDM of electron cloud in atom is exactly opposite in sign to nuclear EDM. 

A quantum-mechanical derivation of this result for an arbitrary non-relativistic system of point-like charged particles with EDMs has been done by Schiff  \cite{Schiff}. Schiff also mentioned that his  theorem is violated by the finite nuclear size. The effect of the finite nuclear size was implemented as the nuclear Schiff moment which was introduced in Refs. \cite{Sandars,Hinds,SFK,FKS1986}. An electrostatic interaction between the nuclear Schiff moment and  electrons produces atomic and molecular EDM. Refs. \cite{Sandars,Hinds} calculated  the finite nuclear size effect of the proton EDM.   Refs. \cite{SFK,FKS1986} calculated (and named) the nuclear Schiff moment  produced by the P,T-odd nuclear forces. It was shown in \cite{SFK} that the contribution of the P,T-odd forces to the nuclear EDM and Schiff moment is $\sim 40$ times larger than the contribution of the nucleon EDM.   An additional 2-3 orders of magnitude enhancement appears in nuclei with the octupole deformation \cite{Auerbach}.

The suppression factor for  the atomic EDM relative to the nuclear EDM, proportional to a very small ratio ($\sim 10^{-9}$) of the squared nuclear radius to the squared atomic radius,  is partly compensated by the factor $Z^2 R_S$, where $Z$ is the nuclear charge and $R_S$ is the relativistic factor \cite{SFK}.  

  The Schiff theorem is also violated by the magnetic interaction \cite{Schiff,Khriplovich}. Corresponding atomic EDMs produced by the nuclear EDM and electron-nucleus magnetic interaction have been calculated in Ref. \cite{FGP}. In light atoms  this mechanism of atomic EDM dominates but in heavy atoms it is smaller than the effect of the finite nuclear size since the latter very rapidly increases with the nuclear charge, as $Z^2 R_S$, while the magnetic effect increases slower, as $Z R_M$ where $R_M$ is the relativistic factor for the magnetic effect \cite{FGP} .
 
There is no complete shielding in ions. For example, in a molecular ion the shielding factor for the nuclear EDM is $(Z_i/Z)(M_n/M_m)$, where $Z_i$ is the ion charge, $Z$ is the nuclear charge, $M_n$ is the nuclear mass and  $M_m$ is the total molecular mass \cite{FlambaumKozlov}. 

Screening of time-dependent  electric  field is incomplete. External electric field can even be enhanced if the frequency of electric filed  oscillations is in resonance with atomic or molecular transition.   Screening of oscillating field has been investigated  in Refs.  \cite{DFSS1986,Flambaum2018,DBGF2018,FlambaumSamsonov,Bao}.


Penetration of electric field to atomic nucleus may  affect nuclear reactions - see e.g.  Refs. \cite{LaserNucl,Ho}, where  neutron capture to a nucleus,  enhanced by a laser field, is discussed. 
 For example, electric field  may admix large   s-wave neutron capture amplitude to a kinematically suppressed p-wave amplitude in a p-wave resonance  and produce effects such as asymmetry  in the photon distribution correlated with the  direction  of the electric field.  This  effect  is somewhat similar to the effects of parity violating interaction which mixes s-wave and p-wave resonances.  Such kinematic enhancement factor (ratio of s-wave and p-wave amplitudes) is $\sim 10^3$ for slow neutrons. Another enhancement factor  is due to  a very small energy interval between the energy levels in a compound nucleus. These two enhancement factors lead to a $10^6$ enhancement of the parity violating effects in neutron reactions predicted in Refs. \cite{SF1980, SF1982, Nucl1,Nucl2} and confirmed in numerous experiments 
 involving a hundred of  p-wave resonances in many nuclei - see  reviews \cite{Gribakin,Bowman}).  The same enhancement may appear if the s and p resonances mixing is produced by the electric field.

The aim of this paper is to show that for atoms and  molecules in a non-stationary state the electric filed on the nucleus does not vanish and to derive formulas for this electric field. This electric field may interact with nuclear EDM and affect nuclear reactions.

\section{ Shielding theory for non-stationary atomic states}



The Hamiltonian of a neutral  atom in an external electric field along the $z$-axis $E^{ext} =E^{ext}_z$ may be presented as 
\begin{eqnarray}\label{HE}
H_E=H_0 - E^{ext}_z D_z \,, \\
D_z=-e \sum_{k=1}^{Z} z_k\,,
\end{eqnarray}
where $H_0$ is the Schrodinger or the Dirac Hamiltonian for the atomic electrons in the absence of the external field $E^{ext}_z$, $Z$ is the number of the electrons and protons,
$-e $ is the electron charge (i.e. $e$ is the proton charge), $z_k$ is the $z$-axis projection of the electron position relative to the nucleus. We assume that the nuclear mass is infinite and neglect very small effects of the   Breit and  magnetic interactions. The total electric field on the nucleus  may be presented as ${\bf E}^t={\bf E}^{ext} +<{\bf E}^e> $, where the $z$ component of the electron electric field on the nucleus is
\begin{equation}\label{Ee}
E^{e}_z= 
e\sum_{k=1}^{Z}\frac{z_k}{r_k^3}=\frac{i}{Z e\hbar}[P_z,H_0] \,,
\end{equation}
where $P_z=\sum_{k=1}^{Z}p_{z,k}$ is the total momentum of the atomic electrons. The second equality follows from the differentiation of the nuclear Coulomb potential in the Dirac or Schrodinger Hamiltonian $H_0$ since the total electron momentum $P_z$ commutes with the electron kinetic energy and the electron-electron interaction.  Similarly, the $z$ component of the total electric field on the nucleus may be presented as 
\begin{equation}\label{Et}
E^{t}_z=\frac{i}{Z e\hbar}[P_z,H_E] \,.
\end{equation}
In agreement with the Schiff theorem, in a  stationary state $| n >$ expectation value of  the total electric field on the nucleus vanishes,  $<n| E^{t}_z | n >=0$,  since $<n | [P_z,H_E] | n> = (\epsilon_n  - \epsilon_n) <n | P_z | n>=0$.  

 A non-stationary state may be presented as a sum over stationary states. For brevity, we include two states in the sum:  
\begin{equation}\label{psi}
\psi=c_a \psi_a \exp{(-\frac{i}{\hbar} \epsilon_a t )}  + c_b \psi_b \exp{(-\frac{i}{\hbar} \epsilon_b t )}
\end{equation}
 In such state the $z$ component of the total electric field on the nucleus is 
\begin{eqnarray}\label{EtP}
\nonumber
<E^{t}_z> = -\frac{i (\epsilon_a- \epsilon_b)}{Ze \hbar} [c_a^* c_b <a|P_z|b>  \exp{(\frac{i(\epsilon_a - \epsilon_b)t}{\hbar}  )}  \\
 -  c_a c_b^* <b|P_z|a>  \exp{(-\frac{i(\epsilon_a - \epsilon_b)t}{\hbar}  )}]\,.\,
\end{eqnarray}
It is also instructive to present $<E^{t}_z>$ using  a substitution of the nonrelativistic expression for the momentum, $P_z=-\frac{i m }{e\hbar} [H_E, D_z]$:
\begin{eqnarray}\label{EtD}
\nonumber
<E^{t}_z> = -\frac{(\epsilon_a - \epsilon_b)^2 m}{Ze^2 \hbar^2} \times\\
\nonumber 
[c_a^* c_b <a|D_z|b>  \exp{(\frac{i(\epsilon_a - \epsilon_b)t}{\hbar}  )}  \\
 + c_a c_b^* <b|D_z|a>  \exp{(-\frac{i(\epsilon_a - \epsilon_b)t}{\hbar}  )}],
\end{eqnarray}
where $m$ in this expression is electron mass. Note that $\psi_a=\psi^{(0)}_a + \delta \psi_a $ and $\psi_b=\psi^{(0)}_b + \delta \psi_b $ are eigenfunctions of the Hamiltonian $H_E$ including interaction with the external electric field $E^{ext}$. Here $\psi^{(0)}_a $ and $\psi^{(0)}_b $ are  eigenfunctions of the unperturbed Hamiltonian $H_0$.

We have two different cases here. If the matrix element between unperturbed wave functions satisfy selection rules for the electric dipole matrix element, i.e.  $<\psi^{(0)}_a | D_z | \psi^{(0)}_b> $ is not equal to zero, we have oscillating  electric  filed on the nucleus  even in the absence of  external electric field $E^{ext}$.  Indeed, atom in such state Eq. (\ref{psi}) has oscillating electric dipole moment $<D_z>$  which produces electric field on the nucleus.
For real $c_a$, $c_b$ and $ <a|D_z|b> $ we obtain
\begin{equation}\label{Ereal}
<D_z> = 2 c_a c_b <a|D_z|b> \cos{(\frac{(\epsilon_a - \epsilon_b)t}{\hbar}  )} 
\end{equation}
and 
\begin{equation}\label{Ereal}
<E^{t}_z> = -  \frac{ 2 (\epsilon_a - \epsilon_b)^2 m}{Ze^2 \hbar^2} c_a c_b <a|D_z|b> \cos{(\frac{(\epsilon_a - \epsilon_b)t}{\hbar}  )}.
\end{equation}
Numerical estimate for the amplitude of this oscillating field is 
\begin{equation}\label{Enumerical}
|E^{t}_z| \sim \frac{  (\epsilon_a - \epsilon_b)^2 }{Z(\text{eV})^2} 10^{7} \text{V}/\text{cm}.
\end{equation}
We assumed $c_ac_b \sim 1$. For $Z \sim 1$ and  $(\epsilon_a - \epsilon_b)$ equal to few eV, this field may  exceed by three orders of magnitude  electric fields,
 which have been used to measure neutron and atomic EDM.
However,  this is a very rapidly oscillating electric field.  In the case of a  small oscillation frequency,   the electric field is strongly suppressed by the factor  $(\epsilon_a - \epsilon_b)^2$. 

When  $<\psi^{(0)}_a | D_z | \psi^{(0)}_b > =0$, we should consider effect produced by  the external electric field $E^{ext}$.  Substitution of the perturbation theory expression for $\delta \psi$ gives, for real $c_a$, $c_b$ and matrix elements of electric dipole moment operator  $ <a|D_z|n> $ and $ <b|D_z|n> $, 
 \begin{equation}\label{Estark}
<E^{t}_z> = \frac{ 2 c_a c_b  (\epsilon_a - \epsilon_b)^2 m}{Ze^2 \hbar^2} \alpha_{a,b} E^{ext}_z \cos{(\frac{(\epsilon_a - \epsilon_b)t}{\hbar}  )},
 \end{equation}
where 
\begin{eqnarray}\label{Stark}
\nonumber
\alpha_{a,b}  = \sum_n \frac{<a|D_z|n><n|D_z|b>}{\epsilon_a - \epsilon_n} + \\
 \frac{<a|D_z|n><n|D_z|b>}{\epsilon_b - \epsilon_n} 
\end{eqnarray}
is the Stark amplitude between the states $a$ and $b$, and $|n>$ are intermediate states in the perturbation theory sum for $\delta \psi$. We see that constant external electric field $E^{ext}_z$ is transformed into oscillating electric field on the nucleus.  Numerical estimate for the amplitude of the field on the nucleus  for  $c_ac_b \sim 1$ is 
\begin{equation}\label{Enumerical1}
|E^{t}_z| \sim \frac{  (\epsilon_a - \epsilon_b)^2 }{Z(27 \text{eV})^2} E^{ext}_z.
\end{equation}
Thus, for  $\epsilon_a - \epsilon_b$ smaller than atomic unit of energy $27$ eV, electric field on the nucleus is smaller than external electric field.

\section{Shielding in non-stationary molecular  states}  

In molecules in a stationary rotational state the screening of external electric field is produced by both electrons and nuclei. However, in a non-stationary state  electric field on the nucleus is proportional to mass of the particles which produce this screening  - see Eq.  (\ref{EtD}). Mass of nuclei is from three  to six  orders of magnitude bigger than mass of electron. Therefore, electric field  on the nucleus may be significantly bigger in molecules compare to atoms (for equal values of $\epsilon_a - \epsilon_b$). Indeed, nuclei in molecules are slow, they  are not as efficient screeners as electrons in the case when (electron) electric field varies. We observed similar enhancement  when considered screening of oscillating external electric field  in molecules   \cite{Bao}.

Let us consider electric field on the  nucleus 1 in a diamagnetic diatomic molecule which is given by the following expression:
\begin{eqnarray}\label{Emol}
{\bf E^{(1)}}= -\frac{i}{Z_1 e\hbar}[{\bf P^{(1)}},H_E] \,,
\end{eqnarray}
where ${\bf P^{(1)}}$ is the momentum of the nucleus 1, $Z_1$ is its charge  and the Hamiltonian $H_E$ includes both nuclei and electrons.  We may subtract from the momentum ${\bf P^{(1)}}$  the contribution of the center of mass motion with velocity ${\bf v}={\bf P}_t/M_t$, where ${\bf P}_t$ is the total momentum and $M_t$ is the total mass of the molecule:
\begin{equation}\label{Pi}
{\bf \Pi}^{1}={\bf P}^{(1)}- M^{(1)} {\bf P}_t / M_t \, .
\end{equation} 
Center of mass momentum commutes with the Hamiltonian, therefore, we may re-write Eq. (\ref{Emol}) using commutator with ${\bf \Pi}^{1}$: 
\begin{eqnarray}\label{EmolPi}
{\bf E^{(1)}}= -\frac{i}{Z_1 e\hbar}[{\bf \Pi}^{1},H_E] \,,
\end{eqnarray}
The expectation value of the total electric field on the  nucleus 1 in the state (\ref{psi})  is 
\begin{eqnarray}\label{EtP}
\nonumber
<{\bf E^{(1)}}> = -\frac{i (\epsilon_a - \epsilon_b)}{Z_1 e \hbar} \times \\
\nonumber
[c_a^* c_b <a|  {\bf \Pi}^{1} |b>  \exp{(\frac{i(\epsilon_a - \epsilon_b)t}{\hbar}  )}  \\
 -  c_a c_b^* <b|{\bf \Pi}^{1} |a>  \exp{(-\frac{i(\epsilon_a - \epsilon_b)t}{\hbar}  )}].\,\,\,\,\,
\end{eqnarray}
Now we can use the following relation from Ref. \cite{Bao}, where we have neglected terms proportional to electron mass in comparison with the terms proportional to the nuclear masses:
\begin{equation}
{\bf \Pi}^{1} = i \mu[{\bf R},H'_E]\,,
\end{equation}
where $\mu=M_1 M_2/(M_1 +M_2)$ is the reduced nuclear mass, ${{\bf R}=\bf R^{(1)}- R^{(2)}}$ is the relative coordinate for  the first and second nucleus and $H'_E$ is the Hamiltonian with subtracted contribution of the center of mass motion. Using this relation we obtain 
  \begin{eqnarray}\label{EtDmol}
\nonumber
<{\bf E^{(1)}}> = -\frac{(\epsilon_a - \epsilon_b)^2  \mu }{Z_1 e \hbar^2} \times \\
\nonumber 
[c_a^* c_b <a| {\bf R} |b>  \exp{(\frac{i(\epsilon_a - \epsilon_b)t}{\hbar}  )}  \\
 + c_a c_b^* <b|  {\bf R} |a>  \exp{(-\frac{i(\epsilon_a - \epsilon_b)t}{\hbar}  )}].
\end{eqnarray}
We consider rotational molecular states $|a>=Y_{00}(\theta,\phi) \Psi$ and $|b>=Y_{10}(\theta,\phi)\Psi$ in a diamagnetic diatomic polar molecule, where $Y_{00}$  and $Y_{10}$ describe rotational $L=0$ and $L=1$ molecular states and $\Psi$ is an internal molecular wave function,  describing  electron state and nuclear vibrational state, which are the same for $<a|$ and $<b|$. Assuming real $c_a$ and $c_b$ we obtain  
\begin{eqnarray}\label{ErealMol}
\nonumber
<E^{(1)}_z> = -\frac{ 2 (\epsilon_a - \epsilon_b)^2 \mu   R }{\sqrt{3}Z_1 e \hbar^2} 
c_a c_b  \cos{(\frac{(\epsilon_a - \epsilon_b)t}{\hbar}  )}, \\
\end{eqnarray}
where $R$ is the distance between the nuclei. Note that the second nucleus experiences  electric field of opposite sign (${\bf R} \to - {\bf R}$). The electric field is inversely proportional to the nuclear charge, therefore, the electric forces are equal in magnitude and have opposite sign,  so there is no acceleration of the center of mass. 

The  oscillation frequency of the electric field on the nucleus may be many  orders of magnitude smaller than frequency of optical transitions in atoms. Indeed, the interval between molecular rotational levels is $\sim \mu /m_e$ times smaller than the interval between the electron levels. 
Moreover, molecules may  have doublets of levels with the energy interval much smaller than the rotational energy. Note however that in this case there is strong  suppression of the field by the small factor  $(\epsilon_a - \epsilon_b)^2$.

\section{Conclusion}

The Schiff's shielding theorem  about total screening of electric field on atomic nucleus  does not apply to atoms and molecules in a non-stationary state, i.e. electric field on a nucleus in a neutral system  in a non-stationary state may be  not equal to zero. The field on the nucleus is much bigger in molecules than in atoms (for equal values of $\epsilon_a - \epsilon_b$) since nuclei in molecules are slow and do not produce such efficient screening in a non-stationary case as electrons do. The field enhancement factor is the ratio of nuclear  mass to electron mass. In principle, electric field penetrating to the nucleus makes nuclear EDM observable and may also affect nuclear reactions. 

Schiff theorem has a simple classical explanation.  Application of a homogeneous electric field to a neutral system does not lead to its motion. In this case the nucleus does not move, despite the fact that  it has non-zero charge. This means that the total electric field on the nucleus is zero. The case of a neutral system in a non-stationary state is different. Nuclei may move, keeping center of mass of the system at rest.      

For brevity, we presented non-stationary state as a sum  of two basis stationary states. In general case the number of terms in the sum may be arbitrary large and we have to add summation over basis states ($\sum_{b>a}$) to the formulas for electric field on the nucleus.

\vspace{3mm}
\textit{Acknowledgements} --- 
The work was supported by the Australian Research Council Grants No.\ DP230101058 and DP200100150.

\end{document}